\documentclass[conference]{IEEEtran}
\IEEEoverridecommandlockouts
\usepackage{amsmath,amssymb,amsfonts}
\usepackage{algorithmic}
\usepackage{graphicx}
\usepackage{textcomp}
\usepackage[dvipsnames]{xcolor}
\usepackage{fontawesome}
\usepackage{bbding}
\usepackage{listings}
\usepackage{array}
\usepackage{colortbl}

\usepackage{url}
\usepackage{tikz}
\usepackage{multirow}
\usepackage{pifont}
\usepackage{mdframed}
\usepackage{comment}
\usepackage{hyperref}

\usetikzlibrary{fit, positioning, arrows, arrows.meta,
  decorations.pathreplacing, decorations.markings, shapes}

\usepackage{xspace}
\usepackage[nolist,nohyperlinks]{acronym}

\newcommand{\toolnametitle}{Vivienne\xspace}
\newcommand{\toolname}{\textsc{Vivienne}\xspace}
\newcommand{\toolnameinv}{\textsc{Vivienne}$_{\text{inv}}$\xspace}
\newcommand{\toolnameunroll}{\textsc{Vivienne}$_{\text{unroll}}$\xspace}
\newcommand{\wasm}{WebAssembly\xspace}

\newcommand{\wrongr}[1]{\textcolor{red}{#1}}

\definecolor{keywordCommentColor}{rgb}{0.090000, 0.55, 0.20}
\definecolor{stringColor}{rgb}{0.558215, 0.000000, 0.135316}
\definecolor{typeColor}{rgb}{0.6, 0.000000, 0.3}
\definecolor{localColor}{rgb}{0.6, 0.000000, 0.3}
\definecolor{ndkeywordColor}{rgb}{0.0, 0.558215, 0.558215} 
\definecolor{commentsColor}{rgb}{0.0, 0.558215, 0.558215} 
\definecolor{keywordColor}{rgb}{0.000000, 0.000000, 0.635294}
\definecolor{newgray}{rgb}{0.3, 0.3, 0.3}
\definecolor{agreen}{rgb}{0.0, 0.36, 0.15}
\definecolor{rgreen}{rgb}{0.13, 0.26, 0.12}

\definecolor{reviewgreen}{rgb}{0.13, 0.50, 0.12}

\newcommand{\review}[1]{#1}

\newcommand{\basicCodeStyle}{\ttfamily\footnotesize\color{newgray}}

\newcommand{\xmark}{\ding{55}\xspace}%
\newcommand{\cmark}{\ding{51}\xspace}%

\lstdefinestyle{cstyle}{%
  language=C,
  frame=tb,
  numberblanklines=false,
  escapeinside=||,
  basicstyle=\basicCodeStyle,
  framextopmargin=1pt,
  numbers=left, xleftmargin=.03\textwidth, %
  keywordstyle=\color{keywordColor}\bfseries,
  ndkeywordstyle=\color{ndkeywordColor}\bfseries,
  identifierstyle=\color{black}\ttfamily,
  commentstyle=\itshape\ttfamily\textcolor{commentsColor},
  stringstyle=\color{stringColor}\ttfamily,
}

\lstdefinelanguage{Wasm}{
  keywords={,module, memory, export, secret, public, 
  const, register, import, loop, load8_s, load8_u,store8, xor,
  symb_exec, param,  eq, load, tee, ne, set, data,elem,
  get_local,get_global,set_local, set_global, tee_local, if, then,
  block, br_if, func, store, add, shl, lt_s, get, type, end, br,
  return, result, sconst, },
  keywordstyle=\color{keywordColor}\bfseries,
  identifierstyle=\color{black}\ttfamily,
  sensitive=false,
  comment = [l]{;},
  morecomment=[s]{(;}{;)},
  commentstyle=\itshape\ttfamily\textcolor{commentsColor},
  morecomment=[s][\color{keywordCommentColor}]{\$\ },
  stringstyle=\color{Mahogany}\ttfamily,
  morestring=[b]",
  morekeywords = [2]{i32,i64},
    keywordstyle = [2]\color{RedViolet}\bfseries, 
  morekeywords = [3]{local, global}, 
  keywordstyle = [3]\color{agreen}\bfseries,
  morekeywords=[4]{ l1,l2},
  keywordstyle=[4]\color{newgray}\bfseries
}

\lstdefinestyle{wasmstyle}{%
  language=wasm,
  frame=tb,
  numbers=left,
  numberblanklines=false,
  escapeinside=||,
  basicstyle=\basicCodeStyle,
  numbers=left, xleftmargin=.03\textwidth, %
  }

\lstdefinestyle{plainwasm}{%
  language=wasm,
  escapeinside=||,
  basicstyle=\basicCodeStyle,
  }

\begin{acronym}
\acro{RelSE}{Relational Symbolic Execution}
\acro{SE}{Symbolic Execution}
\acro{WASI}{\wasm System Interface}
\acro{AST}{Abstract Syntax Tree}
\acro{SMT}{Satisfiability Modulo Theories}
\acro{CT}{Constant Time}
\acro{MAC}{Message Authentication Code}
\acro{JIT}{Just-in-time}
\acro{AES}{Advanced Encryption Standard}
\acro{DES}{Data Encryption Standard}
\end{acronym}

\newcommand{\alt}           {~|~}
\newcommand{\add}       [2] {\texttt{Add}(#1,#2)}
\newcommand{\sub}       [2] {\texttt{Sub}(#1,#2)}
\newcommand{\myle}      [2] {\texttt{Le}(#1,#2)}

\newcommand{\load}      [2] {\texttt{Load}(#1,#2)}
\newcommand{\store}     [3] {\texttt{Store}(#1,#2,#3)}

\newcommand{\loadins}       {\texttt{load}}
\newcommand{\storeins}      {\texttt{store}}
\newcommand{\brif}      [1] {\texttt{br\_if}~{#1}}
\newcommand{\trueval}       {\texttt{true}}
\newcommand{\lspace}        {~~}
\newcommand{\llspace}       {~~~~}
\newcommand{\val}       [1] {\langle #1 \rangle}
\newcommand{\dval}      [2] {\langle #1, #2 \rangle}

\newcommand{\cons}          {\!\!::\!\!}
\newcommand{\state}     [5] {\langle #1, #2, #3, #4, #5 \rangle}

\makeatletter
\def\lst@makecaption{%
  \def\@captype{table}%
  \@makecaption
}
\makeatother

\begin{document}

\title{\toolnametitle: Relational Verification of Cryptographic Implementations in WebAssembly}

\author{\IEEEauthorblockN{Rodothea Myrsini Tsoupidi}
\IEEEauthorblockA{
\textit{KTH Royal Institute of Technology}\\
Stockholm, Sweden \\
tsoupidi@kth.se}
\and
\IEEEauthorblockN{Musard Balliu}
\IEEEauthorblockA{
\textit{KTH Royal Institute of Technology}\\
Stockholm, Sweden \\
musard@kth.se}
\and
\IEEEauthorblockN{Benoit Baudry}
\IEEEauthorblockA{
\textit{KTH Royal Institute of Technology}\\
Stockholm, Sweden \\
baudry@kth.se}
}

\maketitle

\begin{abstract}

We investigate the use of relational symbolic execution to counter timing side channels in \wasm programs.
We design and implement \toolname, an open-source tool to automatically analyze \wasm cryptographic libraries for constant-time violations. Our approach features various optimizations that leverage the structure of WebAssembly and automated theorem provers, including support for loops via relational invariants. 
  We evaluate Vivienne on 57 real-world cryptographic
  implementations, including a previously unverified implementation
  of the HACL* library in \wasm. 
  %
  The results indicate that Vivienne is a practical solution for
  constant-time analysis of cryptographic libraries in \wasm.
\end{abstract}


\section{Introduction}
\label{sec:intro}
%
The introduction of
\wasm~\cite{haas_bringing_2017}, a portable low-level language with
focus on security and efficiency, has led to an array of
security-sensitive applications. Cryptography libraries such as
libsodium~\cite{libsodium} and HACL*~\cite{protzenko_formally_2019}
are a prime example of such applications.
Unfortunately, \wasm programs can be vulnerable to different types of
attacks~\cite{255318}, including timing side channels.
%

The constant-time programming discipline is a well-known practice
to defend against timing attacks~\cite{MolnarPSW05,almeida_verifiable_2016}.
The main idea is to disallow the program's control flow and the memory
access patterns that depend on program secrets.
This is surprisingly challenging because many cryptographic routines
are
human-written~\cite{libsodium,bearssl,stuber_torstenstuebertweetnacl-webassembly_2019-1}
and thus, prone to errors, while compilers that preserve constant time
are yet to emerge~\cite{libsodium,bearssl}.
%
This motivates the need for verification of constant-time
implementations in \wasm.

Drawing on the verification-friendly structure of \wasm, existing solutions
such as CT-wasm~\cite{watt_ct-wasm_2019-1} enrich the  \wasm type system with security annotations to enforce constant time.
The efficiency of CT-wasm comes at the expense of a conservative analysis, e.g., by considering the whole memory as secret, thus leading to false positives or refactoring of constant-time programs.
This paper explores the use of  \ac{RelSE} to verify constant-time implementations in \wasm.
The approach relies on an accurate modelling of the memory and other program optimizations, enabling a precise analysis that scales to real-world cryptographic
implementations.
In summary, this paper offers the following contributions:

\begin{itemize}
\item An \ac{RelSE}-based approach for verifying constant-time
  implementations in \wasm
  programs.
\item An automated invariant generation technique for
  analyzing implementations with loops.
\item A thorough evaluation on 45 secure implementations and 12
  insecure implementations in \wasm, including the previously
  non-verified \wasm implementation of HACL* (WHACL*). 
\item \toolname, an open-source implementation of the
  approach.
\end{itemize}

\section{Problem Setting}
\label{sec:background}

This section presents the problem setting, including the constant-time
policy, and background on \wasm and related works.

\subsection{Constant-time Policy}
\label{ssec:ctime}

Constant-time programming discipline is a software-based
defense against timing side-channel attacks.
This discipline relies on the constant-time
policy~\cite{almeida_verifying_2016}, which classifies values as
secret (\texttt{high}) and public (\texttt{low}).
The policy constrains the control-flow instructions and the memory
operations to solely depend on public values, thus disallowing any
secret-dependent control-flow instructions and memory accesses.
Intuitively, the policy requires that any program executions with the
same \texttt{low} values execute the same instructions and yield the
same memory access patterns, independently of \texttt{high}
values. This indicates that execution time of the program is not
affected by secret data.
%
%
%

\begin{lstlisting}[style=cstyle,
  caption={C function \texttt{tls1\_cbc\_remove\_padding}},
  label=lst:cfunc_loop]
int tls1_cbc_remove_padding(const SSL *s,
	   SSL3_RECORD *rec, unsigned bs,
	   unsigned mac_size) {
 int ii, i, j;
 int l = rec->length;
 ii = i = rec->data[l-1]; /* padding_length */
 i++;
 ...
 for (j=(int)(l-i); j<(int)l; j++)
  if (rec->data[j] != ii) /* Incorrect padding */
   return -1;
 ...
}
\end{lstlisting}

Listing~\ref{lst:cfunc_loop} reports a code snippet of the OpenSSL's
Lucky 13 timing vulnerability~\cite{al_fardan_lucky_2013} to illustrate the issue.
Function \texttt{tls1\_cbc\_remove\_padding} removes the padding from
a decrypted message that contains the plain text (secret), the
\ac{MAC} tag, and the padding.
The size of the padding affects the execution time, which in turn
reveals information about the size of the plain text.
Specifically, \texttt{rec->data} holds the
decrypted message together with the \ac{MAC} tag and the padding, and
is thus secret.
Variables \texttt{i} and \texttt{ii} (line 6) contain the last item of
array \texttt{rec->data}, which holds the padding size.
Hence, the number of iterations of the \texttt{for} loop at line 9
depends on the secret-dependent variable \texttt{i}, which affects the
execution time of the function.
Similarly, the guard of \texttt{if} statement at line 10
depends on  \texttt{ii}, which is also secret.
Memory accesses also reveal information through timing  due to the presence of caches.
At line 10, the access to \texttt{rec->data[j]} reveals
information about the value of index \texttt{j} by timing its presence in the cache. 



\subsection{\wasm}
\label{ssec:wasm}

\wasm~\cite{haas_bringing_2017} is a stack-based typed low-level language
serving as backend for both client-side computations, e.g., web browsers, and  server-side computations~\cite{255318} including
stand-alone applications~\cite{clark2019standardizing}.
With some exceptions~\cite{stuber_torstenstuebertweetnacl-webassembly_2019-1}, \wasm code is compiler generated, e.g., via
LLVM  with support for C, C++,
and Rust.  Other languages, like Python and Julia, also provide support for \wasm.
WASI Libc~\cite{clark2019standardizing} is
a library built on top of WASI system calls to enable I/O and memory management for \wasm programs.

The execution model of \wasm~\cite{haas_bringing_2017}
consists of 1) an execution stack $es$ that stores
the instructions; 2) a value stack $vs$ that holds the input
arguments of the instructions, 3) a linear memory, and
4) the local and the global stores.
\wasm has a structured control flow; for indirect calls
(\texttt{call\_indirect}), the call destination is an index to a
function table; for conditional branch (\texttt{br\_if}), the branch
destination is an index $i$ to enter (\texttt{loop}) or exit
(\texttt{block}) the $i$th scope.
Memory operations read from (\texttt{load}) and write to (\texttt{store})
the linear memory, and global variables are visible to all functions in a module. A function may also define local variables \texttt{lv$n$} including the function parameters.
Modules are collections of functions with their own linear memory,
and global variables~\cite{haas_bringing_2017}.

Listing~\ref{lst:wfunc_loop} shows an example \wasm module.
The code is a simplified compiled version (using clang-10) of the C
code in Listing~\ref{lst:cfunc_loop}.
The code consists of a module (line 1-33), which imports a memory
instance (\texttt{"\_memory"}) from another module \texttt{\$env}
(line 3) and declares function \texttt{tls1\_cbc\_remove\_padding}
(line 4).
The function takes four input parameters of type 32-bit integer and
returns a 32-bit value (line 5).
At line 6, the function declares five local variables and the rest of
the function consists of the function body.
The  block at line 8 performs multiple
initializations before the beginning of the loop (line 15).
At line 10, instruction \texttt{local.tee} stores the top value of
$vs$ (here \texttt{rec->data + 1}) to \texttt{lv6} and
pushes the same value back to $vs$. 
%
At line 15, the loop starts by loading \texttt{lv6} and \texttt{lv1}
to $vs$.
Instruction \texttt{i32.add} adds these two values and pushes back the
result to $vs$.
Finally, instruction \texttt{i32.load8\_u} loads from the linear memory (\texttt{"\_memory"}) the value at the index taken from
the top of $vs$, i.e.\ the result of the addition.
The loop body executes until instruction \texttt{br\_if},
which reads one value from $vs$; if the value is non zero
(\texttt{true}), the execution breaks out of the outermost block
(lines 8-31), whereas if the value is zero (\texttt{false}), 
the execution continues to the next instruction, \texttt{br}, which
unconditionally jumps back to the beginning of the loop (line 15).

\begin{lstlisting}[style=wasmstyle,
  caption={Wasm function \texttt{tls1\_cbc\_remove\_padding}},
  label=lst:wfunc_loop]
(module
  ...
  (import "env" "_memory" (memory (;0;) 2))
  (func $tls1_cbc_remove_padding (type 2)
	(param i32 i32 i32 i32) (result i32)
    (local i32 i32 i32 i32 i32)
    ...
    block  ;; label = @1
      ...
      local.tee 6 ;; tee (rec->data + l)
      ...
      local.tee 1 ;; tee (1 - ii)
      ...
      block  ;; label = @2
        loop  ;; label = @3
          local.get 6 ;; get l
          local.get 1 ;; get (j - l)
          i32.add
          i32.load8_u ;; load data[j] from memory
	  ...
	  i32.const 1
	  local.set 4 ;; store return value
	  ...
          local.set 1 ;; j++
	  ...
          br_if 2 (;@1;) ;; break if j >= l
          br 0 (;@3;) ;; continue to loop
        end
      end
      ...
    end
    ...))
\end{lstlisting}

\wasm programs may be vulnerable to timing side-channel attacks.  The
constant-time policy for \wasm concerns control-flow instructions, i.e.\
\texttt{br\_if}, \texttt{if},
\texttt{br\_table}, and \texttt{call\_indirect}, and the memory
operations, i.e.\ \texttt{load} and \texttt{store}.

\subsection{Related Work}
\label{ssec:relwork}

Several works have aimed at improving the security of
\wasm~\cite{255318,watt_ct-wasm_2019-1,
  watt_weakening_2019,vassena_automatically_2020,narayan_swivel_2021}.
CT-wasm~\cite{watt_ct-wasm_2019-1} proposes a type system to check the
constant-time policy.
Type checking is very efficient but it suffers from the annotation
burden and the conservative nature of the analysis.
In CT-wasm, this is reflected by the treatment of the whole memory as
secret, e.g.\ requiring that every \texttt{load} operation returns a
\texttt{high} value, which may require refactoring of the programs to
make them amenable to the analysis (e.g., \texttt{poly1305\_blocks}
and \texttt{poly1305\_update} functions of a \wasm TweetNaCl
implementation~\cite{stuber_torstenstuebertweetnacl-webassembly_2019-1}).
%
%
Our approach aims at overcoming these limitations by means of
\ac{RelSE}, using a more accurate memory model and no extensive
annotation burden. 
\review{ Moreover, we expect our analysis to yield less false positives because
it relies on symbolic execution which is more precise than security type systems.
For example, an expression such as $\texttt{secret} - \texttt{secret}$
would be correctly identified as the constant $\texttt{0}$.  }
However, as we will see, our solution comes with a computation cost
due to the increased precision. 

Almeida et al.~\cite{almeida_verifying_2016} use product programs to
verify constant-time for C implementations.
A drawback of verifying the constant-time policy for
high-level languages is that the analysis does not provide guarantees
on the security of the generated code (see \texttt{ct\_select}
implementations~\cite{daniel_binsecrel_2020}).
Daniel et al.~\cite{daniel_binsecrel_2020} verify constant-time
programs at the binary level using \ac{RelSE}.
Web browsers using \wasm typically leverage \ac{JIT} compilation,
which does not result in binary file generation.
Moreover, the verification of constant-time at the \wasm level
provides opportunities for optimization due to \wasm's structured
design.
HACL*~\cite{zinzindohoue_hacl_2017} uses a high-level specification
language to generate a formally verified cryptographic library that is
available in different languages including C and
\wasm~\cite{protzenko_formally_2019}.

\section{\toolname: \ac{RelSE} for \wasm}
\label{sec:tool}



\begin{figure}
  \centering
  \begin{tikzpicture}[
  scale=1., transform shape,
  tool/.style={font=\normalsize, minimum width = 1cm,
    thick, minimum height = 0.8cm, rounded corners=2pt},
  result/.style={circle, draw, thin, minimum width = 0.8cm},
]

    \node[tool, draw,
      align = center] at (0,0) (d0) {Wasm\\Modules};

  \node[tool,
    draw,
    below=0.4cm of d0,
    align = center] (pol) {Security\\Policy};

  \node[tool,
    draw,
    below= 0.4cm of pol,
    align = center] (script) {Entry\\Point};

  \node[tool, draw,
    above right = 10pt and 90pt of pol,
    anchor= center,
    minimum width = 2cm,
    minimum height = 1.2cm, 
    align = center ] (relse) {Relational Symbolic\\Execution};

  \draw[->, >={Latex}] (d0.east) -- (relse.west);
  \draw[->, >={Latex}] (pol.east) -- (relse.west);
  \draw[->, >={Latex}] (script.east) -- (relse.west);

  \node[tool, draw,
    below right=35pt and 18pt of relse.center,
    minimum height = 20pt,
    text width= 40pt,
    align = center,
    dotted] (invar) {Invariant Generation};


  \node[tool, draw,
    below left=1.2cm and 0.3cm of relse.center,
    minimum height = 1.2cm,
    text width = 2cm,
    align = center ] (formula) {Formula Simplification};

  \node[draw, thick, inner xsep=1em, inner ysep=1em, fit=(relse) (invar) (formula),
    label={[fill=white,yshift=-0.3cm]\toolname}] (tool) {};

  \draw[->, >={Latex}] (relse.206) -- node[font=\footnotesize,left]{(1)$e$}(relse.206 |- formula.north);
  \draw[->, >={Latex}] (formula.60) -- node[font=\footnotesize,right,align=center]{(2)sat/\\unsat}
  (formula.60 |- relse.south);
  
  \draw[->, >={Latex}] (invar.170) -- node[font=\footnotesize,above]{(1)$e$}
  (formula.east |- invar.170);
  \draw[->, >={Latex}] (formula.-5) -- node[font=\footnotesize,below, align=center]{(2)sat/\\unsat}
  (invar.west |- formula.-5);

  \draw[->, >={Latex}] (relse.-30) -- node[font=\footnotesize, left]{(1)$loc$}(relse.-30 |- invar.north);
  \draw[->, >={Latex}] (invar.105) -- node[font=\footnotesize, right]{(2)$I$}(invar.105 |- relse.south);

  \node[tool, draw,
    below= 0.7cm of formula,
    minimum height = 10pt,
    align = center ] (solver) {SMT Solver};

  \draw[->, dotted, >={Latex}] (formula.-115) -- node[font=\footnotesize,
    below left = 0pt and 2pt,
    inner sep = 0pt,
    fill=white]{(1)$\phi$}(formula.-115 |- solver.north);

  \draw[->, dotted, >={Latex}] (solver.80) -- node[font=\footnotesize,below right,
    inner sep=0pt,
    align=center,
    fill=white]{(2)sat/unsat}(solver.80 |- formula.south);

  \node[below right= 12pt and 35pt of relse, anchor=west] (bug) {\faBug};
  \node[below left= -12pt and -5pt of bug.center] (bug2) {\faBug};
  \node[above right= -12pt and -5pt of bug.center] (bug3) {\faBug};
  \draw[->, >={Latex}] (relse) -- ([xshift=20pt]relse.east) |- (bug);
  
  \node[right = 35pt of relse, anchor=west, align=center] (ok) {Verify\\ CT\Checkmark};
  \draw[->, >={Latex}] (relse) -- (ok);
  
\end{tikzpicture}
  \caption{\label{fig:arch} \toolname Architecture}
\end{figure}

\toolname analyzes \wasm implementations with respect to constant
time.
%
Figure~\ref{fig:arch} shows a high-level view of the tool.
\toolname takes three inputs: 1) the \textit{\wasm modules} containing
the functions to analyze, 2) the \textit{security policy} annotating the
memory regions and the parameters of the entry function,
and 3) the \textit{entry point} describing the entry
function to analyze.
Then, \toolname performs \ac{RelSE} on the entry function, reporting the
 discovered constant-time
vulnerabilities (if any).  We describe the
different components of \toolname using Listing~\ref{lst:wfunc_loop} as a running example. 

\textbf{\wasm Modules}
The  modules include the entry function to verify and its
dependencies, possibly involving different modules.
For example, the module in Listing~\ref{lst:wfunc_loop} imports the
memory from another module \texttt{\$env} (line 3) and defines
function \texttt{tls1\_cbc\_remove\_padding} (lines 4-28).

\textbf{Security Policy and Entry Point}
The security policy specifies the parts of the memory and the 
arguments of the entry function that contain public or secret
values.
Listing~\ref{lst:policyentry} reports the policy for function
\texttt{tls1\_cbc\_remove\_padding}.
%
%
%
The policy specifies the bytes 2000 to 2039 (i.e.\ pointer \texttt{s})
and the memory of struct \texttt{rec} as public (not shown),
and the bytes 2048 to 2111 (i.e.\ \texttt{rec->data}) as
secret, thus reflecting the specification in
Listing~\ref{lst:cfunc_loop}.
Moreover, \toolname requires the 
code of the modules (line
8) and the \textit{entry function} (lines 9-11).
The latter includes the security policy for its arguments
which can be either concrete or symbolic values.
Lines 9--11 specify the concrete and symbolic arguments for 
analyzing function \texttt{tls1\_cbc\_remove\_padding} via
\ac{RelSE}.
The function takes four arguments: 1) the memory index of \texttt{s}; 
2) the memory index of struct 
\texttt{rec}; 3) the block size, which is a public symbolic
value; and 4) the \texttt{\ac{MAC}} size which is also a
public symbolic value.
\toolname recognizes public (secret) symbolic values
that start with letter \texttt{l} (\texttt{h}).

\begin{lstlisting}[style=wasmstyle,
    caption={Security policy and Entry Function},
    label=lst:policyentry]
(module $env
 (memory (;0;) $memory (export "_memory") 2)
 (public (i32.const 2000) (i32.const 2039));;s
 ...
 (secret (i32.const 2048) (i32.const 2111));;data 
)
;;definition of tls1_cbc_remove_padding-Listing 2
...
(symb_exec "tls1_cbc_remove_padding"
  (i32.sconst 2000) (i32.sconst 2040) ;; concrete
  (i32.sconst l|$_1$|)   (i32.sconst |l$_2$|))  ;; symbolic
\end{lstlisting}




\begin{figure}
\[\arraycolsep=1.6pt
\begin{array}{lcl}
  v~(values)            &::=& h_{n} \alt l_{n} \alt c,   \llspace  c\in\mathbb{Z}, n\in\mathbb{N}_0   \\
  \rho~(relational\ values) &::=& \langle v, v \rangle                                              \\
  e~(expressions)       &::=& \rho \alt \add{e}{e} \alt \sub{e}{e} \alt ...                         \\
                        &   &  |~\myle{e}{e} \alt \load{e}{\mu}                                     \\
  i~(instructions)      &::=& \brif{l} \alt ... \alt \loadins, \lspace l\in\mathbb{N}_0             \\
  \hline
  \mu~(memory)          &::=& \bot \alt \store{e}{e}{\mu}                                           \\
  st~(stack)            &::=& \varnothing \alt e :: st                                              \\
  pc~(path\ condition)  &::=& \review{\trueval} \alt e \land pc                                     \\
  es~(execution\ stack) &::=& \varnothing \alt i::es                                                \\
  lv~(local\ variables) &::=& \{lv_0 \mapsto e, ..., lv_n \mapsto e\}                               \\
\end{array}
\]
\caption{\label{fig:sym} Symbolic Data Structures}
\end{figure}

\textbf{\acl{RelSE}}
\toolname uses the above-mentioned inputs to initiate 
\ac{RelSE}~\cite{farina_relational_2019-1} for the
entry function.
\ac{RelSE} performs symbolic execution on relational states representing
two program executions with identical public values but
different secret values.
%
We now describe the ingredients underpinning the constant-time analysis with \toolname.  



\paragraph{Symbolic State}
A symbolic state $\sigma$ consists of 1) the
execution stack $es$, that contains the \wasm instructions,
2) the symbolic stack $st$, 3) the symbolic memory $\mu$, 4) the
symbolic local (and global) variables $lv$, and 5) the path
condition $pc$.
Figure~\ref{fig:sym} summarizes these five components of a symbolic state $\sigma =
\state{ es}{st}{\mu}{lv}{pc}$.
By convention, the values starting with $h$ ($l$) are
secret (public). Our symbolic analysis operates on pairs of symbolic values $\rho$.
We write $\rho_{|l}$ ($\rho_{|r}$) to denote the first (second) element of a pair $\rho$.
For public values, we have that $\rho_{|l} = \rho_{|r}$ and 
write $\val{v}$, while for secret values $\rho_{|l}$ and $\rho_{|r}$ may differ.
We lift this notation to expressions and
the memory as expected.

\paragraph{Execution Path Exploration}
We use small-step symbolic evaluation to analyze the instructions.
At every step, the analysis takes a symbolic state as input and
returns a list of symbolic states that correspond to the feasible
execution paths.
We visit the instructions in a depth-first search fashion and collect all path conditions $pc$ to check  path feasibility using an \ac{SMT} solver.
%

\paragraph{Symbolic Stack}
The symbolic stack  holds symbolic expressions $e$ resulting from
stack operations on symbolic values.
Consider the \texttt{get} instructions at lines 16--17 in Listing~\ref{lst:wfunc_loop} with 
the current symbolic memory $\mu$ and empty symbolic stack $st$.
The program loads the symbolic expressions of \texttt{lv6}
i.e.\ $\val {2112}$,  and \texttt{lv1}
i.e.\ $\sub{\val{1}}{\load{\val{2111}}{\mu}}$ to the stack $st$.
At line 18, the analysis of instruction \texttt{add} pops the two symbolic expressions off the
stack $st$ and pushes back the result,
$\add{\val{2112}}{\sub{\val{1}}{\load{\val{2111}}{\mu}}}$.

\paragraph{Memory Operations}
\label{par:mo}
%
When analyzing a memory operation at index $e$, as in \ $\state{
  \loadins\cons es}{e\cons st}{\mu}{lv}{pc}$ or $\state{
  \storeins\cons es}{e_1\cons e\cons st}{\mu}{lv}{pc}$, the analysis
generates a formula, $\phi = (T(e)|_{|r} \neq T(e)_{|l})$ to check
that the index is not secret-dependent.
The function $T: e \to \langle Exp, Exp\rangle$ translates the index
expression $e$ to a pair of \ac{SMT} expressions $ Exp$.
If $e$ only depends on public values, then for all valuations of $e$,
$e_{|r} = e_{|l}$, thus $\phi$ is \textit{unsatisfiable} and the
memory operation is \texttt{safe}.
However, if $\phi$ is \textit{satisfiable}, then there are concrete
values, such that the memory addresses for the two executions,
$e_{|r}$ and $e_{|l}$, are different.
This is only possible if expression $e$ depends on secret values, and,
thus, the solution to $\phi$ reveals a violation of constant time. 
In our example in Listing~\ref{lst:wfunc_loop}, load operation
\texttt{load8\_u} at line 19 has as index the top value of $st$,
$\add{\val{2112}}{\sub{\val{1}}{\load{\val{2111}}{\mu}}}$.
The policy in Listing~\ref{lst:policyentry} specifies
$\load{\val{2111}}{\mu}$ as secret, i.e.\ $\load{\val{2111}}{\mu} =
\dval{h_1}{h'_1}$ with $h_1 \ne h'_1$.
Thus, the generated formula  $\phi = (2112 + (1 - {h_1})) \neq
(2112 + (1 - {h'_1}))$ is satisfiable for different values of
$h_1$ and $h'_1$.
This means that there exist the two concrete executions
that differ with regards to the memory index, which violates 
constant-time.

\paragraph{Control-flow Instructions}
Like memory operations, control-flow instructions require checking
that boolean expression $e$, as in $\state{ \brif{0}\cons es}{e\cons
  st}{\mu}{lv}{pc}$, is not secret-dependent. Our analysis generates a
formula to check whether the two paths of the relational state take
different branches.
\wasm considers value \textit{zero} as
false and \textit{any non-zero} value as true, hence 
the generated formula is $\phi = (T(e)_{|r} = 0)
\land (T(e)_{|l} \neq 0)$.
Formula $\phi$ is satisfiable only if there is a valuation of $e$ such
that the two executions follow different execution paths, indicating a
violation of the constant-time policy.
%

\textbf{Formula Simplification}
\label{sssec:fs}
When \ac{RelSE} needs to check the constant-time policy for an
expression $e$, it first passes $e$ to the simplification step
(SS).
SS translates the expression to a pair of \ac{SMT}
expressions, $e' = T(e)$, using the theory of bitvectors and arrays
(32-bit indexed byte array), \texttt{QF\_ABV}.
The transformation includes simplification and memoization
steps to reduce the recalculation overhead.
Finally, based on the type of the query, namely memory operation or
control-flow statement, this step generates formula $\phi$.
For our previous example, SS first
translates expression $e =
\add{\val{2112}}{\sub{\val{1}}{\load{\val{2111}}{\mu}}}$ to two SMT
expressions $2112 + (1 - h_1)$ and $2112 + (1 - h'_1)$, which
are then simplified to $2113 - h_1$ and $2113 - h'_1$,
hence the final formula becomes $\phi = (2113 - h_1) \neq (2113 - h'_1)$.
%
%
To solve the simplified formula, \toolname invokes an
\ac{SMT} solver.
For simple formulas, however, the resulting $\phi$ may already be a concrete boolean, e.g., \texttt{false}, allowing \toolname skip a call to the \ac{SMT} solver.



\textbf{\acs{SMT} Solver}
\label{sssec:smtsolver}
After the simplification step, \toolname invokes an \ac{SMT} solver
for solving the simplified formula, $\phi$.
The \ac{SMT} solver of \toolname has two modes, one for small formulas
and one for large and complex formulas.
For small formulas, \toolname uses a solver that provides bindings to
the implementation language of \toolname and thus, has a reduced
communication cost.
However, for larger formulas, the communication overhead is less
significant compared to the benefit of using a more powerful \ac{SMT}
solver.
In particular, for larger queries \toolname uses a portfolio solver
were many solvers take as input the same formula and the solver that
finishes first returns the result.
To decide over which solver mode to use, \toolname uses the
\textit{number of expressions} in the formula.

\textbf{Invariant Generation}
\label{sssec:inv}
\toolname has an optional invariant generation step for analyzing  loops.
When invariant generation is enabled and the analysis visits a
loop at location $loc$, \toolname starts a preprocessing step to
automatically generate a relational invariant $I$.
The invariant defines the variables (local variables, global
variables, and memory) that are public, i.e.\ $I = \{ \forall
x \in V_p\subseteq V.~ x_{|l} = x_{|r} \}$, where $V$ is the set of all
variables modified in the loop and $V_p$ is the subset of the modified
variables that are public.
To discover whether a variable is public or secret, the preprocessing
step queries the \ac{SMT} solver about the security policies of the
modified variables, $V$, \review{after symbolically executing one loop iteration}.
\review{That is, given a variable $x \in V$, the preprocessing step
  generates a query, $\phi = (x_{|l} \ne x_{|r})$.
  If the query is unsatisfiable, then the variable is assumed to be
  public and $x$ is added to $V_p$, otherwise, it is assumed to
  be secret.
  In the special case of $x_{|l} = x_{|r} = c \in \mathbb{Z}$, the
  analysis assumes that $x$ has a symbolic value $c$ and adds the equality constraint $x = c$ to the
  invariant, $I$.}
After generating invariant $I$, the analysis continues with verifying
this invariant.
To do that, \toolname 1) \review{generates fresh symbolic variables} (havoc) for
all modified variables $x\in V$, 2) assumes that the invariant, $I$,
holds, 3) performs \ac{RelSE} on the loop body with the havoced values
and discovers possible vulnerabilities, 4) verifies that the invariant
holds by asserting $I$ on the new relational state.
\review{If the generated invariant is not a loop invariant,
  then the last step will fail.
}
After analyzing the loop body, the analysis continues outside the loop.
\review{The invariant verification algorithm is a generalization of standard (functional) invariant checking, hence we expect the loop analysis to be sound, as supported by the experiments.}

Consider the loop at $loc=15$ in Listing~\ref{lst:wfunc_loop}.
Local variables 1 and 4 are modified in the loop body,
i.e.\ $V=\{lv1,lv4\}$.
Of these, $lv1$ stores \texttt{j} (line 24), which is secret because it
depends on \texttt{rec->data[l-1]} and $lv4$ stores value 1, which
is public.
Thus, $V_p = \{lv4\}$, hence the invariant is $I = \{ lv4_{|l} =
lv4_{|r} \}$.
To analyze the loop, \toolname 1) havocs $lv1$ and $lv4$, 2) assumes
the invariant $I$, i.e.\ that $lv4$ is initially public, 3) performs
\ac{RelSE} at the loop body to discover constant-time vulnerabilities,
and 4) asserts the invariant $I$.
Here, the program assigns $lv4$ only once in the loop body, at line
22, where, $lv4$ takes value one, which is public, and thus,
the invariant $I$ holds.

\textbf{Output} \toolname outputs the discovered constant-time
violations (\faBug), if any, as well as the \ac{SMT} solver-generated
counterexamples that witness these violations.


\toolname is implemented as an extension of the \wasm reference
interpreter~\cite{ref-interpreter} in OCaml, using OCaml compiler 4.06.
\toolname uses the OCaml interface of z3~\cite{de_moura_z3_2008-1} to
generate and simplify the constant-time formulas, and solve queries that
have a small number of expressions.
For larger formulas, \toolname uses a portfolio solver consisting of
four solvers, i.e.\ Boolector~\cite{brummayer_boolector_2009},
Yices2~\cite{dutertre_yices_2014}, CVC4~\cite{barrett_cvc4_2011}, and
Z3~\cite{de_moura_z3_2008-1} running in parallel.
\toolname is publicly available online at
\url{https://github.com/romits800/Vivienne}.

\section{Evaluation}
\label{sec:eval}


%

We evaluate \toolname with respect to three research questions:

\textbf{RQ1: Can we use \ac{RelSE} for
  constant-time analysis of real-world cryptographic implementations
  in \wasm?}
To investigate the effectiveness and efficiency of \ac{RelSE} for
constant-time analysis on \wasm programs, we use \toolname to analyze
the implementations of seven cryptographic libraries within a time
limit of 90 minutes.
%

\textbf{RQ2. To what extent do the automatically generated loop
  invariants affect the scalability and precision  of \ac{RelSE}?}
We evaluate \toolname's support for automatic invariant generation on
our benchmarks and compare it to the results of RQ1.

\textbf{RQ3. How does \toolname compare to existing approaches for
  constant-time analysis of \wasm?}
We compare \toolname with CT-wasm~\cite{watt_ct-wasm_2019-1}  with regards to
simplicity, permissiveness, and efficiency.

%

\subsection{Experimental Setup and Overview of Benchmarks}
\label{ssec:es}
We run the experiments on a machine running Debian GNU/Linux 10 (buster) on an Intel
Core\texttrademark i9-9920X processor 3.50GHz  with 64GB of RAM.
We used the LLVM-10 compiler with WASI
libc~\cite{clark2019standardizing} and two optimization levels
(\texttt{-O0} and \texttt{-O3}) for compiling our C benchmarks to
\wasm.
\toolname uses a time limit of 90 minutes for each benchmark and a
threshold of 1500 expressions to trigger a call to the portfolio
solver.
%

We evaluate \toolname with seven cryptography libraries, including
both constant-time and non-constant-time implementations.
Some benchmarks have been used in prior works
~\cite{watt_ct-wasm_2019-1,daniel_binsecrel_2020} to evaluate
constant-time policies, which provides us with common ground for
comparison.
We extract the security policies for the first two libraries from the
type annotations of CT-wasm~\cite{ctwasm} and use the policies of
Binsec/Rel~\cite{binsec} for the other libraries.
The full details of our benchmarks are available at
\url{https://github.com/romits800/Vivienne_eval}.

\textbf{CT-wasm benchmarks (CTw):} Three handwritten \wasm benchmarks
from CT-wasm~\cite{watt_ct-wasm_2019-1}. We verify the
\texttt{encrypt} and \texttt{decrypt} functions of \texttt{Salsa20}
and \texttt{TEA}, and the \texttt{transform} and \texttt{update}
functions of \texttt{SHA256}.

\textbf{TweetNaCl \wasm (Tw):} \wasm implementation of
TweetNaCl~\cite{stuber_torstenstuebertweetnacl-webassembly_2019-1}  previously verified by
CT-wasm~\cite{watt_ct-wasm_2019-1}.
We verify 
\texttt{core\_hsalsa20}, \texttt{core\_salsa20}, and
\texttt{crypto\_onetimeauth}.

\textbf{WHACL* (WH):}  A formally verified cryptography library compiled to \wasm ~\cite{protzenko_formally_2019}.
%
%
We verify \texttt{Chacha20}, \texttt{Curve25519\_51},
\texttt{Poly1305\_32}, \texttt{Salsa20}, and \texttt{Hash\_SHA2} in WHACL*
 v3.0.0.
To our best knowledge, this is the first time WHACL*  is verified.

\textbf{Libsodium (L0, L3):} A
cryptography library written in C~\cite{libsodium}.
%
\toolname verifies the constant-time implementations of
\texttt{crypto\_aead}, \texttt{crypto\_auth}, \texttt{crypto\_stream},
\texttt{crypto\_onetimeauth}, \texttt{crypto\_core}, and
\texttt{crypto\_hash} for Libsodium  v.1.0.18-stable with
optimization levels \texttt{-O0} and \texttt{-O3}.

\textbf{BearSSL (B0, B3):} An
implementation of SSL/TLS in C.
%
%
We verify the  constant-time functions \texttt{aes\_ct\_cbcenc} and
\texttt{des\_ct\_cbcenc} and the non constant-time functions \texttt{aes\_big\_cbcenc} and \texttt{des\_tab\_cbcenc}. B0 includes the functions with optimization level \texttt{-O0} and B3 is optimization \texttt{-O3}. 

\textbf{Almeida et al.~\cite{almeida_verifying_2016} (A0, A3):}
Five constant-time and three non-constant-time implementations of
\texttt{select} and \texttt{sort}. 
%
We analyze \wasm binaries compiled with optimization
levels \texttt{-O0} and \texttt{-O3}.

\textbf{Lucky 13 (Lu0, Lu3):} A known timing
vulnerability~\cite{al_fardan_lucky_2013} of TLS implementations (see
Listing~\ref{lst:cfunc_loop}). 
%
We analyze function \texttt{tls1\_cbc\_remove\_padding} of
OpenSSL
1.0.1 \cite{repo1}
with optimization levels \texttt{-O0} and \texttt{-O3}.

\begin{table}
  \centering
  
\begin{tabular}{lr|rrrr|rrrr}
  \hline
  \multicolumn{2}{c|}{Bench.}&\multicolumn{4}{c|}{\toolnameunroll}&\multicolumn{4}{c}{\toolnameinv}\\
  \hline
  \texttt{BS}&\texttt{A}&\cmark&\xmark&\texttt{\#FS}&\texttt{\#SS}&\cmark&\xmark&\texttt{\#FS}&\texttt{\#SS}\\
  \hline\hline
  CTw  & 6               &6/6  &0/0  &4K     &0           &6/6   &0/0                       &814      &412\\
  \hline
  Tw   & 3               &3/3  &0/0  &181    &0           &3/3   &0/0                       &320      &164\\ 
  \hline
  \rowcolor{white!80!black}
  WH   & 6               &\wrongr{5}/6  &0/0  &126K   &0  &6/6   &0/0                       &70K      &7K\\
  \hline
  B0  & 4                &2/2  &2/2  &32K    &40          &\wrongr{1}/2   &\wrongr{0}/2     &10K       &873\\
  B3  & 4                &2/2  &2/2  &2K     &40          &\wrongr{0}/2   &\wrongr{1}/2     &158K     &3K\\
  \hline
  L0  & 8                &8/8  &0/0  &113K   &18          &\wrongr{2}/8   &0/0              &21K      &347\\
  L3  & 8                &8/8  &0/0  &9K     &18          &\wrongr{3}/8   &0/0              &3K       &309\\
  \hline
  A0  & 8                &5/5  &3/3  &683    &31          &\multicolumn{4}{c}{\multirow{2}*{No loops}}\\
  A3  & 8                &5/5  &3/3  &55     &9           &\\
  \hline
  \rowcolor{white!80!black}
  Lu0  & 1               &0/0  &{\color{red}0}/1  &25K     &4K         &0/0   &1/1     &539      &217\\
  Lu3  & 1               &0/0  &1/1  &3K     &3K          &0/0   &\wrongr{0}/1     &94        &63\\
  \hline\hline
Sum   & 57               &44/45 &11/12 &-      &-           &21/35  &2/6     &-        &-\\
 \hline
\end{tabular}

  \caption{\label{tab:eval} Verifying 57 cryptography functions with
    \toolname, with unrolling and with \review{invariant inference}.
    The numbers in {\color{red}red} denote incomplete results.}
\end{table}

\subsection{Results}
\label{ssec:res}
This section discusses the evaluation results for each of the
research questions.
Table~\ref{tab:eval} presents the aggregated results of the analysis
with \toolname.
The columns under \texttt{Bench} describe the benchmarks, i.e.\ the
abbreviated library name, \texttt{BS}, and the number of analyzed
algorithms, \texttt{A}.
The next two columns present \toolname's results with loop unrolling
(\toolnameunroll) and with loop invariant (\toolnameinv).
We report the number of verified constant-time implementations,
\cmark, the number of vulnerable implementations \xmark, the number of
formulas subject to simplification, \texttt{\#FS}, and the number of
queries that \toolname propagates to the \ac{SMT} solver,
\texttt{\#SS}.
Note that \texttt{\#SS} is the subset of \texttt{\#FS} that requires a
call to the \ac{SMT} solver.
We highlight in \textcolor{red}{red} the incomplete results.
For example, \toolname with loop unrolling (\toolnameunroll) was able
to verify successfully five out of six implementations of WH within
the time limit of 90 minutes.
%
Appendix~\ref{sec:fulleval} includes the full evaluation results
for \toolnameunroll, while the results for \toolnameinv are available
as supplementary material online~\cite{tsoupidisupp2021}.

\subsubsection{RQ1: Can we use \ac{RelSE} for
  constant-time analysis of real-world cryptographic implementations
  in \wasm? }
To evaluate the effectiveness of \toolname in analyzing cryptographic
libraries, we consider the rate of successfully analyzed algorithms for both secure (\cmark) and insecure (\xmark) implementations.
%
%
%
The summarized results (Sum) in Table~\ref{tab:eval} show that
\toolnameunroll analyzes successfully 44 out of 45 constant-time
implementations and 11 out of 12 non-constant-time implementations for
a total 55/57 implementations.
This corresponds to 96\% success rate while reporting no false
positive.
The two outliers are \texttt{Hacl\_Curve25519\_51\_scalarmult} of WH
and \texttt{tls1\_cbc\_remove\_padding} of Lu0.
The former contains a
loop with 256 iterations, each  generating 9108 queries.
One of these queries affects an increasingly large part of the total
execution time for an iteration.
The corresponding formula models the satisfiability of a branch condition
that depends on the stack pointer, which WHACL* stores in memory.
As a result, the formula has to encode the whole memory, which contributes with 3054 new memory stores for every iteration, thus increasing the time for
the generation and simplification of the formula.
This can be inferred from the results of
Table~\ref{tab:eval}, where the total six implementations of WH
generate 126K formulas (\texttt{\#FS}), of which 80896 correspond to
\texttt{Hacl\_Curve25519\_51\_scalarmult}.

The second outlier is
\texttt{tls1\_cbc\_remove\_padding} with \texttt{-O0}, which contains
a loop with non-constant bound, as reported in line 9 in Listing~\ref{lst:cfunc_loop}. 
The lack of a constant bound forces \toolnameunroll to consider all possible values for \texttt{rec->data[l-1]}, which is an eight-byte value.
This leads to maximum 256 iterations for every path that visits the
loop.
%
%
We find that the optimization level \texttt{-O0} includes a number of
stack operations that modify the memory at every iteration.
As we can see in Table~\ref{tab:eval}, this leads to 25K \texttt{\#FS}
and 4K \texttt{\#SS}.
The former requires on average 0.01 seconds (4 minutes in total) for simplification, whereas the latter requires 0.87 seconds (58 minutes in total) for SMT solving.
%

%
In summary, our results show that \ac{RelSE} can be used to analyze
real-world cryptographic implementations, while the memory operations
and loops remain the main bottleneck for the SMT solver.
\toolnameinv addresses the challenge of loops by generating relational
loop invariants automatically.
Further discussion about the SMT solver results of our
analysis are available as supplementary
material~\cite{tsoupidisupp2021}.

\subsubsection{RQ2. To what extent do the automatically generated loop
  invariants affect the scalability and precision  of \ac{RelSE}?}
Our results in Table~\ref{tab:eval} show that \toolnameinv is able to
successfully analyze constant-time implementations for the first three
benchmark libraries.
It also analyzes successfully the implementations of  WH and Lu0 
that \toolnameunroll could not handle.
%
%
%
Perhaps surprisingly, \toolnameinv performs poorly on the benchmarks
B0, B3, L0, and L3, analyzing only 29\% of the implementations.
%
The main reason is that the havocing of modified variables during the
invariant generation replaces constant values with unbounded symbolic
values.
This triggers a path explosion whenever a conditional instruction is
analyzed with the new symbolic values.
Moreover, it increases the search space for the solver and the
complexity of queries whenever a symbolic value indexes the memory in
store operations.
%
%
In Table~\ref{tab:eval}, the number of solver queries, \texttt{\#SS},
for \toolnameinv is larger than for \toolnameunroll, which reflects the
increase in the complexity because the solver queries
(\texttt{\#SS}) report the formulas that cannot be resolved during
the simplification stage.
For the benchmarks Tw and
B3, the number of queries, \texttt{\#FS}, also increases due to path
explosion.
By contrast, for the benchmarks that \toolnameinv analyzes
successfully, \texttt{\#FS} decreases due to the reduction of loop
iterations by the loop invariant.

In summary, \toolnameinv analyzes
successfully 56\% of the implementations, including  
two implementations for which \toolnameunroll failed.
This shows that \toolnameinv complements \toolnameunroll for
constant-time analysis.

\subsubsection{RQ3: How does \toolname compare to existing approaches for constant-time analysis of \wasm?}
To our best knowledge, CT-wasm~\cite{watt_ct-wasm_2019-1} is the only
constant-time analysis tool for \wasm.
We consider three
dimensions for comparison: 1) simplicity, 2) permissiveness, and 3) efficiency.
Simplicity refers to the required user effort to verifying a
target implementation.
CT-wasm relies on type annotations for the program,
which can be partially inferred~\cite{watt_ct-wasm_2019-1}.
By contrast, \toolname requires only the security policies and
entry-point function, otherwise no further modifications to the
generated \wasm binary are needed.
This reduces the user effort for analyzing a program.
Permissiveness refers to the ability of the method
to analyze and successfully verify cryptographic implementations.
CT-wasm considers the whole memory as secret, which rules out any
secure programs that store public values in memory.
For example, CT-wasm required refactoring three functions of the
TweetNaCl~\cite{stuber_torstenstuebertweetnacl-webassembly_2019-1}
library, i.e.\ \texttt{poly1305\_blocks}, \texttt{poly1305\_update},
and \texttt{poly1305\_finish}, to make it amenable to verification.
By contrast, \toolname  analyzes and verifies
the whole implementation of (\texttt{crypto\_onetimeauth}),
with no modifications to the original code.
%
Moreover, \toolname could analyze and verify 57 \wasm implementations,
including the two libraries CTw and Tw which were verified by
CT-wasm~\cite{watt_ct-wasm_2019-1}.
%
With regards to efficiency, CT-wasm is clearly superior to \toolname
because it relies on type checking, while \toolname performs expensive
symbolic analysis and constraint solving.
However, as we have seen in RQ1, \toolname was still able to analyze
real-world \wasm implementations within a reasonable time limit.

To summarize, \toolname verifies a larger number of cryptographic
implementations than CT-wasm with no need for refactoring and with
minimal annotation efforts at the expense of an efficiency cost.

\subsubsection{Discussion}
\review{Ideally, an accurate analysis should be implemented as close to the
hardware as possible to avoid vulnerabilities introduced by compiler
transformations.
For \toolname, the structured control flow of \wasm facilitates the
analysis, while binary-level analyses face challenges with
unstructured control flow and diversity of
architectures~\cite{balliu_automating_2014,daniel_binsecrel_2020}.
This raises the question of whether constant-time programs at \wasm
level preserve the property at the machine level.
%

%
The machine code generated from a \wasm binary relies on the
compiler of the respective runtime system.
Unfortunately, a direct analysis of this machine code with  tools like
Binsec/Rel~\cite{daniel_binsecrel_2020} is not possible due to the
different calling conventions and implementation details of
Binsec/Rel.
A comparison of \toolname's results at the \wasm level with Binsec/Rel's results at the machine level for the benchmarks L0, L3, B0, B3, A0, A3, Lu0, and Lu3
in Table~\ref{tab:eval} shows that both tools yield the same result
on all benchmarks, except of the \texttt{select} implementations of
 the benchmarks of Almeida et al.~\cite{almeida_verifying_2016}.
The difference is manifested in the compilation of the \texttt{select}
implementations at optimization level \texttt{-O3}, which Binsec/Rel
identifies as insecure.
In our experiments, LLVM-10 with flag \texttt{-O3} compiles all the C
implementations of \texttt{select} (A3 in Table~\ref{tab:eval}) to
one \wasm \texttt{select} instruction.
The compilation from \wasm to machine code translates the \wasm
\texttt{select} instruction either to a constant-time conditional
assignment (safe), e.g.\ \texttt{cmov} for x86, or to a set of
instructions that include a branch instruction (unsafe), depending on the
target machine and the compiler implementation.
%
%
To account for these differences, \toolname provides a command-line
option for treating the \wasm \texttt{select} instruction as unsafe.}

\section{Conclusion}
\label{sec:conc_fw}
This paper presented \toolname, an open-source tool for analyzing
constant-time for \wasm programs.
\toolname relies on \ac{RelSE} and leverages the structure of \wasm to
implement several optimizations, including automated invariant
generation.
We used \toolname to analyze successfully 57 cryptographic
implementations with minimal annotation overhead and no code
refactoring.
Moreover, \toolname is the first tool to verify constant time for the
\wasm implementation of HACL*.

%

 \section*{Acknowledgments}
 We thank anonymous reviewers for their helpful feedback. This work is partially supported by the Wallenberg AI, Autonomous Systems, and Software Program (WASP) funded by Knut and Alice Wallenberg Foundation, the TrustFull project funded by the Swedish Foundation for Strategic Research (SSF), the JointForce project funded by the Swedish Research Council (VR), and Digital Futures.

\bibliographystyle{IEEEtran} \bibliography{bibliography}

\appendices

\section{Evaluation Results}
\label{sec:fulleval}

Table~\ref{tab:unroll_full} shows the complete results of the
evaluation for \toolnameunroll.
The experiments use a time limit of 90 minutes and the reported time
values are in seconds and consist of the average and standard
deviation after five runs.
The first column shows the file name followed by the function that
corresponds to the entry point for the analysis.
Column \texttt{LoC} shows the number of \wasm instructions that the
analysis accesses, column \texttt{AN time} is the analysis time
in seconds.
When \texttt{AN time} is -1, then \toolname was not able to
successfully analyze the respective implementations. 
Column \faBug~shows the number of discovered timing vulnerabilities.
\#FS is the number of formulas during the analysis and  next column
shows the time in seconds for the simplification step.
\#SS is the number of formulas that \toolname forwards to the SMT
solver, followed by the average number of expressions in each formula,
\#Exprs, and the solving time \texttt{SS time}.
\#Exprs is the value that decides selecting the \textit{bindings}
solver or the \textit{portfolio} solver.
In these experiments, for \#Expr $\le$ 1500, \toolname uses the
\textit{bindings} solver, otherwise the portfolio solver.

For example, the first entry for Libsodium -O0 in Table~\ref{tab:unroll_full}
shows the results for the analysis of functio
\texttt{crypto\_aead\_chacha20poly1305\_encrypt} from libsodium
\texttt{aead} module.
\toolname goes through 7720 different \wasm  instructions, not
including the multiple accesses for loops.
The analysis time is $7.89 \pm 0.09$ seconds and the analysis did not
discover any timing vulnerabilities, generated 11507 formulas that took
$0.03 \pm 0.48$ seconds to simplify.
The high value ($0.48$) of the variance depends on the difference in
the complexity of these 11507 formulas.
Of the 11507 formulas, 16 where forwarded to the SMT portfolio solver,
whereas the rest were simple enough for the analysis to infer their
result.
The average number of expressions in these 16 formulas is four and the
average solving time was 0.04 seconds.

\begin{table*}
  \centering
  \begin{tabular}{l|rrrrrrrr}
    \hline
bench/function&LoC&AN time&\faBug&\#FS&FS time&\#SS&\#Exprs&SS time\\
\hline
\multicolumn{9}{c}{CT-wasm}\\
\hline
salsa20/decrypt&515&0.09 $\pm$ 0.00&0&602&$<0.01$&0&&\\
salsa20/encrypt&512&0.10 $\pm$ 0.01&0&602&$<0.01$&0&&\\
sha256/transform&372&0.05 $\pm$ 0.01&0&926&$<0.01$&0&&\\
sha256/update&409&0.18 $\pm$ 0.01&0&1312&$<0.01$&0&&\\
tea/decrypt&80&$<0.01$&0&72&$<0.01$&0&&\\
tea/encrypt&80&0.01 $\pm$ 0.00&0&72&$<0.01$&0&&\\
\hline
\multicolumn{9}{c}{TweetNaCl}\\
\hline
core\_hsalsa20/core\_hsalsa20&356&$<0.01$&0&46&$<0.01$&0&&\\
core\_salsa20/core\_salsa20&412&0.01 $\pm$ 0.00&0&54&$<0.01$&0&&\\
poly1305/crypto\_onetimeauth&787&0.11 $\pm$ 0.00&0&81&$<0.01$&0&&\\
\hline
\multicolumn{9}{c}{WHACL*}\\
\hline
chacha20/Hacl\_Chacha20\_chacha20\_encrypt&1777&669.91 $\pm$ 3.53&0&9665&0.07 $\pm$ 2.77&0&&\\
curve25519\_51/Hacl\_Curve25519\_51\_scalarmult&-1&-1&0&80896&0.07 $\pm$ 0.26&0&&\\
poly1305/Hacl\_Poly1305\_32\_poly1305\_mac&1440&1.34 $\pm$ 0.01&0&829&$<0.01$&0&&\\
salsa20/Hacl\_Salsa20\_salsa20\_encrypt&1887&162.86 $\pm$ 1.56&0&8596&0.02 $\pm$ 0.71&0&&\\
sha256/Hacl\_Hash\_SHA2\_hash\_256&1147&1323.51 $\pm$ 7.13&0&14512&0.09 $\pm$ 4.56&0&&\\
sha512/Hacl\_Hash\_SHA2\_hash\_512&1550&456.20 $\pm$ 4.14&0&12287&0.04 $\pm$ 1.62&0&&\\
\hline
\multicolumn{9}{c}{BearSSL -O0}\\
\hline
aes\_big/br\_aes\_big\_cbcenc\_run&2089&13.04 $\pm$ 0.11&32&1111&$<0.01$&32&3711&0.36 $\pm$ 0.37\\
aes\_ct/br\_aes\_ct\_cbcenc\_run&4857&46.54 $\pm$ 0.76&0&4233&0.01 $\pm$ 0.13&0&&\\
des\_ct/br\_des\_ct\_cbcenc\_run&3841&1560.52 $\pm$ 6.80&0&23463&0.07 $\pm$ 1.23&0&&\\
des\_tab/br\_des\_tab\_cbcenc\_run&1920&24.94 $\pm$ 0.16&8&3301&0.01 $\pm$ 0.05&8&262&$<0.01$\\
\hline
\multicolumn{9}{c}{BearSSL -O3}\\
\hline
aes\_big/br\_aes\_big\_cbcenc\_run&791&7.89 $\pm$ 0.09&32&218&$<0.01$&32&3327&0.22 $\pm$ 0.22\\
aes\_ct/br\_aes\_ct\_cbcenc\_run&1717&1.69 $\pm$ 0.01&0&493&$<0.01$&0&&\\
des\_ct/br\_des\_ct\_cbcenc\_run&993&6.49 $\pm$ 0.03&0&952&0.01 $\pm$ 0.19&0&&\\
des\_tab/br\_des\_tab\_cbcenc\_run&581&3.20 $\pm$ 0.03&8&381&0.01 $\pm$ 0.15&8&262&$<0.01$\\
\hline
\multicolumn{9}{c}{Libsodium -O0}\\
\hline
aead/crypto\_aead\_chacha20poly1305\_encrypt&7720&369.83 $\pm$ 1.33&0&11507&0.03 $\pm$ 0.48&16&4&0.04 $\pm$ 0.00\\
auth/crypto\_auth\_hmacsha256&13913&4856.64 $\pm$ 27.94&0&47679&0.10 $\pm$ 0.52&0&&\\
chacha20/crypto\_stream\_chacha20&3313&228.04 $\pm$ 1.61&0&8756&0.03 $\pm$ 0.51&2&4&0.04 $\pm$ 0.00\\
poly1305/crypto\_onetimeauth\_poly1305\_donna&3685&20.78 $\pm$ 0.09&0&1671&0.01 $\pm$ 0.07&0&&\\
salsa20/crypto\_core\_salsa20&1628&11.99 $\pm$ 0.04&0&3513&$<0.01$&0&&\\
sha256/SHA256\_Transform&11692&136.11 $\pm$ 0.95&0&8299&0.02 $\pm$ 0.06&0&&\\
sha256/crypto\_hash\_sha256&13225&536.25 $\pm$ 3.84&0&18712&0.03 $\pm$ 0.11&0&&\\
sha512/crypto\_hash\_sha512&13351&295.80 $\pm$ 3.18&0&12993&0.02 $\pm$ 0.08&0&&\\
\hline
\multicolumn{9}{c}{Libsodium -O3}\\
\hline
aead/crypto\_aead\_chacha20poly1305\_encrypt&1971&45.06 $\pm$ 0.29&0&896&0.05 $\pm$ 0.67&16&4&0.04 $\pm$ 0.00\\
auth/crypto\_auth\_hmacsha256&3256&562.00 $\pm$ 4.19&0&4559&0.12 $\pm$ 5.32&0&&\\
chacha20/crypto\_stream\_chacha20&956&0.29 $\pm$ 0.01&0&253&$<0.01$&2&4&0.04 $\pm$ 0.00\\
poly1305/crypto\_onetimeauth\_poly1305\_donna&940&11.20 $\pm$ 0.07&0&223&0.05 $\pm$ 0.58&0&&\\
salsa20/crypto\_core\_salsa20&483&0.01 $\pm$ 0.00&0&52&$<0.01$&0&&\\
sha256/SHA256\_Transform&2171&0.01 $\pm$ 0.00&0&479&$<0.01$&0&&\\
sha256/crypto\_hash\_sha256&2980&28.06 $\pm$ 0.66&0&1643&0.02 $\pm$ 0.66&0&&\\
sha512/crypto\_hash\_sha512&2844&6.20 $\pm$ 0.06&0&1344&$<0.01$&0&&\\
\hline
\multicolumn{9}{c}{Almeida -O0}\\
\hline
naive\_select/ct\_select\_u32\_naive&49&0.03 $\pm$ 0.00&1&9&$<0.01$&3&15&$<0.01$\\
select\_v1/ct\_select\_u32\_v1&149&$<0.01$&0&14&$<0.01$&0&&\\
select\_v2/ct\_select\_u32\_v2&93&$<0.01$&0&10&$<0.01$&0&&\\
select\_v3/ct\_select\_u32\_v3&70&$<0.01$&0&9&$<0.01$&0&&\\
select\_v4/ct\_select\_u32\_v4&70&$<0.01$&0&9&$<0.01$&0&&\\
sort/sort3&254&0.18 $\pm$ 0.00&1&298&$<0.01$&14&68&$<0.01$\\
sort\_multiplex/sort3\_multiplex&276&0.02 $\pm$ 0.00&0&89&$<0.01$&0&&\\
sort\_negative/sort3\_negative&209&0.16 $\pm$ 0.01&1&245&$<0.01$&14&68&$<0.01$\\
\hline
\multicolumn{9}{c}{Almeida -O3}\\
\hline
naive\_select/ct\_select\_u32\_naive&5&$<0.01$&0&0&&0&&\\
select\_v1/ct\_select\_u32\_v1&5&$<0.01$&0&0&&0&&\\
select\_v2/ct\_select\_u32\_v2&5&$<0.01$&0&0&&0&&\\
select\_v3/ct\_select\_u32\_v3&5&$<0.01$&0&0&&0&&\\
select\_v4/ct\_select\_u32\_v4&5&$<0.01$&0&0&&0&&\\
sort/sort3&84&0.07 $\pm$ 0.01&3&21&$<0.01$&3&229&0.02 $\pm$ 0.01\\
sort\_multiplex/sort3\_multiplex&74&0.10 $\pm$ 0.00&3&17&$<0.01$&3&229&0.02 $\pm$ 0.02\\
sort\_negative/sort3\_negative&74&0.09 $\pm$ 0.01&3&17&$<0.01$&3&229&0.02 $\pm$ 0.02\\
\hline
\multicolumn{9}{c}{lucky13 -O0}\\
\hline
tls1\_cbc\_remove\_padding\_lucky13/tls1\_...\_lucky13&-1&-1&5&24978&0.01 $\pm$ 0.06&4027&35698&0.87 $\pm$ 0.60\\
\hline
\multicolumn{9}{c}{lucky13 -O3}\\
\hline
tls1\_cbc\_remove\_padding\_lucky13/tls1\_...\_lucky13&133&960.17 $\pm$ 15.52&5&3144&$<0.01$&3106&3080&0.25 $\pm$ 1.03\\
\hline

  \hline
  \end{tabular}
  \caption{\label{tab:unroll_full} Evaluation results with \toolnameunroll}
\end{table*}

\end{document}